\documentclass[12pt,preprint]{aastex}

\usepackage{graphicx}

\def\beq{\begin{equation}}
\def\eeq{\end{equation}}
\def\bea{\begin{eqnarray}}
\def\eea{\end{eqnarray}}

\def\eg{{\it e.g.,~}}
\def\ie{{\it i.e.,~}}

\begin{document}

\title{Non-Gaussian Posteriors arising from Marginal Detections}

\author{Bruce A. Bassett}
\affil{South African Astronomical Observatory, Observatory, Cape Town, South Africa}
\affil{Department of Mathematics and Applied Mathematics, University of Cape Town, Rondebosch, 7700, Cape Town, South Africa}
\affil{African Institute for Mathematical Sciences, Muizenberg, South Africa}
\affil{Centre for High Performance Computing, 15 Lower Hope St, Rosebank, Cape Town, South Africa}
\affil{Perimeter Institute
for Theoretical Physics, 31 Caroline St. N., Waterloo, ON, N2L 2Y5,Canada}

\author{Niayesh Afshordi}
\affil{Perimeter Institute
for Theoretical Physics, 31 Caroline St. N., Waterloo, ON, N2L 2Y5,Canada}
\affil{Department of Physics and Astronomy, University of Waterloo, 200 University Avenue West, Waterloo, ON, N2L 3G1, Canada }


\begin{abstract}
We show that in cases of marginal detections ($\sim 3\sigma$),
such as that of Baryonic Acoustic Oscillations (BAO) in
cosmology,  the often-used Gaussian approximation to the 
full likelihood is very poor, especially beyond $(\sim 3\sigma$).
This can radically alter confidence intervals on
parameters and implies that one cannot naively extrapolate
$1\sigma$ error bars to $3\sigma$ and beyond. 
We propose a simple fitting formula which corrects for
this effect in posterior probabilities arising from marginal
detections. Alternatively the full likelihood should be used for
parameter estimation rather than the Gaussian approximation of a 
just mean and an error. 
\end{abstract}
\maketitle
Making observational detections in cosmology is hard. Two to three sigma evidence for new physics is common in the literature and 
the community is rightly skeptical of results with marginal statistical significance. Even in the case of $5\sigma$ results, detections can be questioned if the results are surprising or conflicting data is released. 

When detections are expected there is a natural tendency to accept them more readily. Baryon Acoustic Oscillations (BAO) \citep[BAO; \eg][]{bao_review} are a good example. The original BAO results of the SDSS and 2dF teams \citep{sdss_bao, 2df_bao} were at less than $3\sigma$ significance. Nevertheless, the detection has been essentially unanimously accepted by the community \citep[although see][]{LVBL} despite the difficulty of localizing the BAO peak, \eg illustrated by the shift in BAO results between 2007 and 2009 in \citet{will} and \citet{dr7} and recent studies of mock catalogs \citep{sim} which suggest that the BAO peak would be invisible in at least $10\%$ of SDSS DR7-sized samples.  

Such detections are easy to accept since the detected peak is precisely in the place where it was expected to be. This willingness to accept marginal detections carries two dangers. First, the signal may actually be pure statistical fluctuation and hence provide precise but inaccurate knowledge. Second, it can actively discourage publication of other studies which are apparently at odds with the `detection'. 

\begin{figure}
\includegraphics[width=\linewidth]{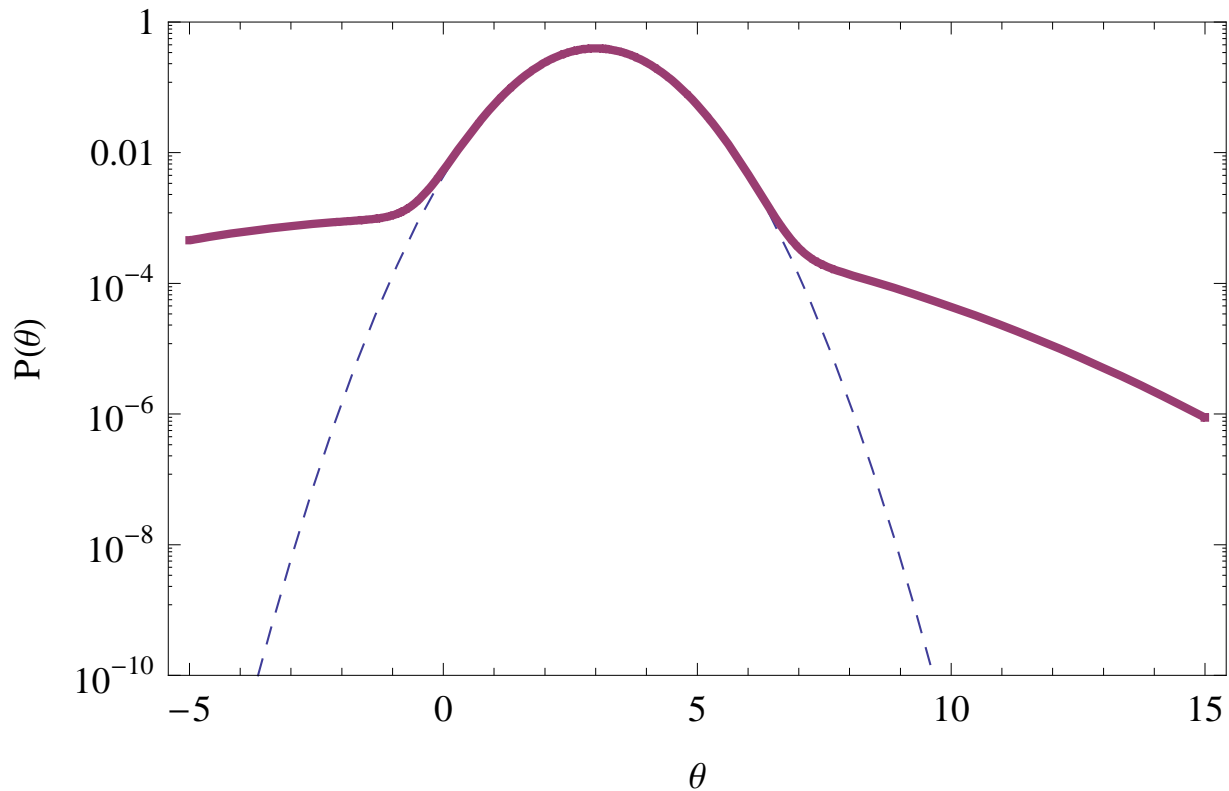}
\caption{An example of how the posterior probability distribution has extended non-Gaussian wings, for finite detection probability: A peak is detected at 99\% confidence level ($P_{detect} = 0.99$) at $\theta = 3 \pm 1$, while the gaussian prior of $\theta = 0 \pm 4$ was assumed. The dashed curve shows the naive gaussian posterior for $\theta$, which {\it e.g.,} would suggest $\theta =7$ is ruled out at $4\sigma$ level, while it is actually allowed at 99.5\% confidence level.}
 \label{BAO_pdf}
\end{figure}

In this {\em letter}, we point out that the possibility that the detection is not real, but just a statistical fluctuation due to noise, has a significant impact on use of the data for parameter estimation. This arises because the full posterior typically has very non-gaussian tails even if the likelihood is gaussian, as illustrated in Figure (\ref{BAO_pdf}). Like a scorpion, the sting is in the tail.

The prototypical example we have in mind is the use of the BAO peak for cosmology, but the principle is valid generally. While the current best BAO combined detection is at the $\sim 4.9 \sigma$ level from the SDSS, 6df and WiggleZ catalogues \citep{blake_wigglez}, lower significance detections will always mark the frontier of the subject as we push to higher redshifts and different samples.The search for BAO with the SDSS, 2QZ and 2SLAQ survey quasars illustrates our point. While they show a peak in the expected place ($105 h^{-1}$Mpc, they also show a (presumably) fake peak near  $85 h^{-1}$Mpc \citep{Sawangwit}. A more standard example is the recent 6df survey which found  $2.4 \sigma$ evidence for BAO while the WiggleZ detection is $3.2 \sigma$ at $z = 0.6$. This will continue with a number of BAO first detections still to come in the next few years:
\begin{itemize}
\item The first separate detections of the radial and transverse BAO peaks which will yield $H(z)$ and $d_A(z)$
separately. There is a claim of detection of the radial BAO \citep{radial}. However this is somewhat controversial and the associated uncertainty illustrates the main points of this paper \citep{crit_rad,crit_rad2}.
\item The first detection of BAO in photometric redshift surveys. \footnote{Current surveys such as Mega-Z lack the number density to reduce shot noise to a level where detection is possible.  DES \citep{des} and PanSTARRS \citep{ps1} should provide the first detections while LSST will provide exquisite results \citep{LSST}.}
\item The first detection of BAO in cluster data. The current status is the 2-2.5$\sigma$ evidence from the maxBCG cluster catalogue \citep{maxbcg}.
\item The first detection of BAO in neutral hydrogen, HI. 
\item The first detection of BAO in the Lyman-$\alpha$ forest \citep{ly-BAO}. This is a method that will be employed by both the BOSS and LAMOST surveys.
\item The first $z > 2$ BAO detection with Lyman break galaxies.
\end{itemize}
In addition,  as future surveys progress, it will be tempting to split a given volume up into narrower redshift bins to provide more data points. In doing so it is standard to ignore the statistical significance of the BAO detections in forecasting the resulting constraints on dark energy, just as it has been standard to ignore them in using current BAO results: prominent examples include the WMAP7 analysis \citep{Komatsu:2010fb}, the SDSS supernova survey \citep{kessler} and the grid marginalisation component of the Supernova Legacy (SNLS) 3-year ana	lysis \citep{snls}, all of which use the gaussian approximation to SDSS BAO results results.

As a toy model to illustrate the impact of ignoring the finite detection probability of the peak, consider  the posterior probability derived from a BAO experiment, $P(\theta | d, detected)$ where $d$ is the data, $\theta$ are the parameters to be estimated (\eg $w_0, w_a, \Omega_k$) and {\it detected} signifies the assumption that the apparent peak is not just noise, \ie the underlying model is correct. Then the full posterior is \citep{press,BEAMS}:
\begin{equation}
P(\theta | d) = P_{detect} P(\theta | d, detected) + (1- P_{detect}) P(\theta),
\end{equation}
where $P_{detect}$ is the statistical significance of the detection and $P(\theta)$ is the prior probability for the parameters which coincides with $P(\theta | d, noise)$, the knowledge gained if the apparent peak is assumed to be pure noise, since in this case there is no new information.  One of the most important implications of this formula is that the resulting posterior is very non-gaussian, with catastrophic widening where the likelihood drops below the prior. An example of this non-gaussian distribution is shown in Figure \ref{BAO_pdf}. An important component of the full posterior is that it must be correctly normalized: $\int P(\theta | d) d\theta = 1$.

\begin{figure}
\includegraphics[width=0.9\linewidth]{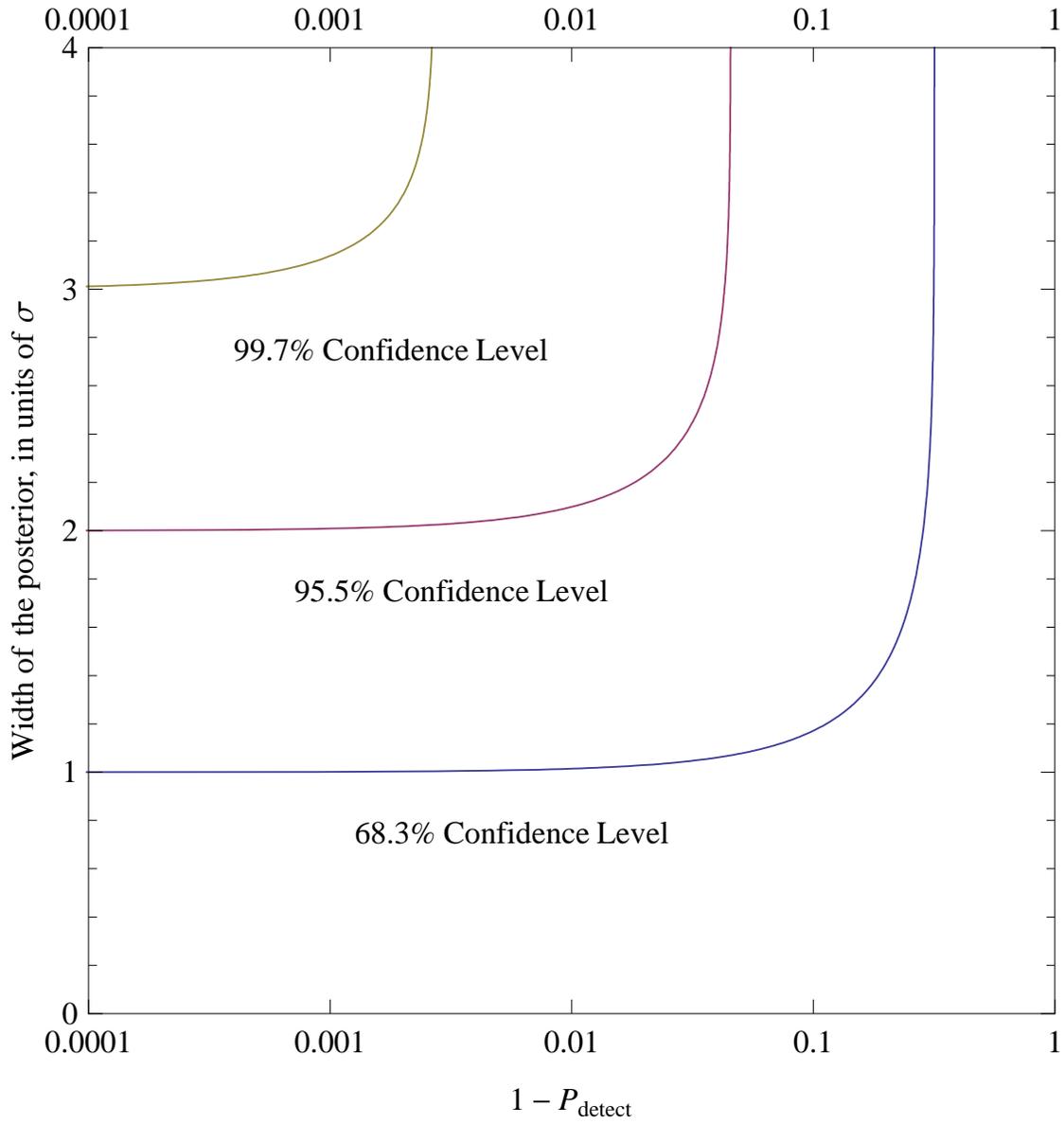}
 \caption{This figure shows how $68.3\%$, $95.5\%$, and $99.7\%$ confidence levels for the width of the posterior probability distribution (associated with $1$, $2$, and $3\sigma$ gaussian errors) rapidly expand with decreasing detection probability, $P_{detect}$.}
 \label{BAO_sigma}
\end{figure}

Assuming that the prior is much wider than the measurement, \ie $P(\theta | d, detected) \gg P(\theta)$ near the peak of the likelihood, Figure (\ref{BAO_sigma}) shows how the errorbars grow as one decreases the detection probability $ P_{detect}$. This is done by finding the $68.3\%$, $95.5\%$, and $99.7\%$ regions, using the normalization $P_{detect} < 1$ for a gaussian distribution.

Let us look at the example of $P_{detect} = 0.99$ (Figure \ref{BAO_pdf}). Although the $1$ and $2-\sigma$ error bars are essentially unaffected, these are of little interest in parameter estimation since any real conclusions must be supported at the 99.7$\%$ ($3\sigma$) level or better. At this level, Figure (\ref{BAO_sigma}) shows that there is a dramatic change: there are {\em no} constraints on the parameter at all at this significance, despite the likelihood $P(\theta | d, detected)$ possibly claiming wonderful constraints.
What is happening is that one is transitioning from the likelihood to the prior when one moves sufficiently far from the maximum likelihood value such that:
\begin{equation}
P(\theta | d, detected) = \frac{ (1-P_{detect})}{ P_{detect}} P(\theta) \simeq  (1- P_{detect}) P(\theta).
\end{equation}
Since the prior should vary with $\theta$ much less rapidly than the likelihood, there is a point at which one's constraints are actually driven by the prior, not the data. The basic concept is simple: one cannot extrapolate data beyond its realm of validity, as illustrated by the figures. 

\begin{figure*}
\includegraphics[width=0.47\linewidth]{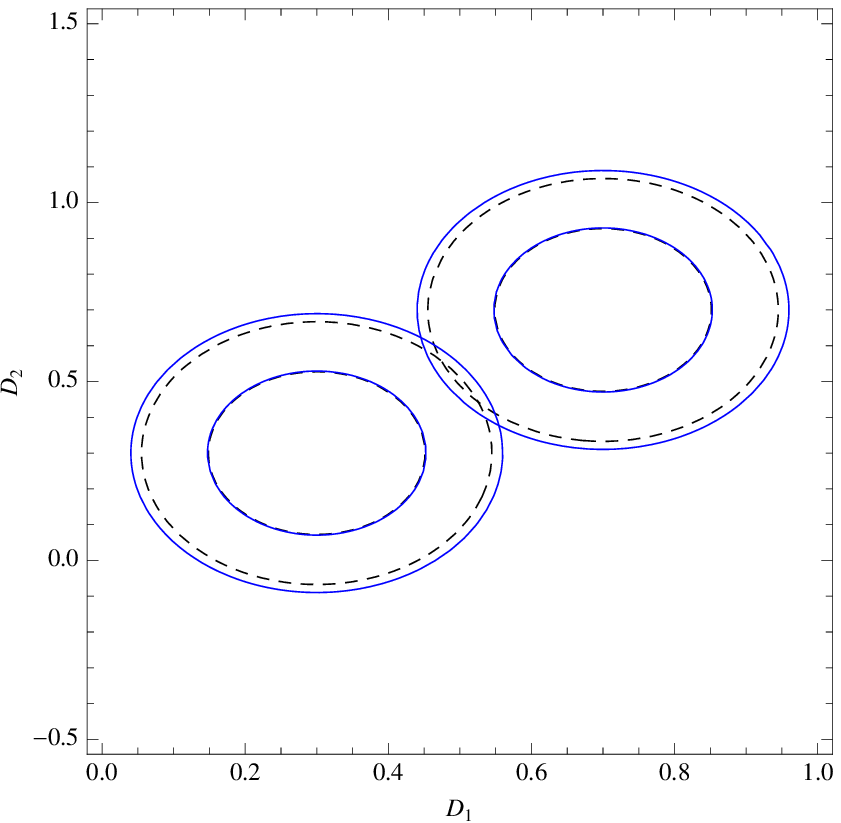}
\includegraphics[width=0.47\linewidth]{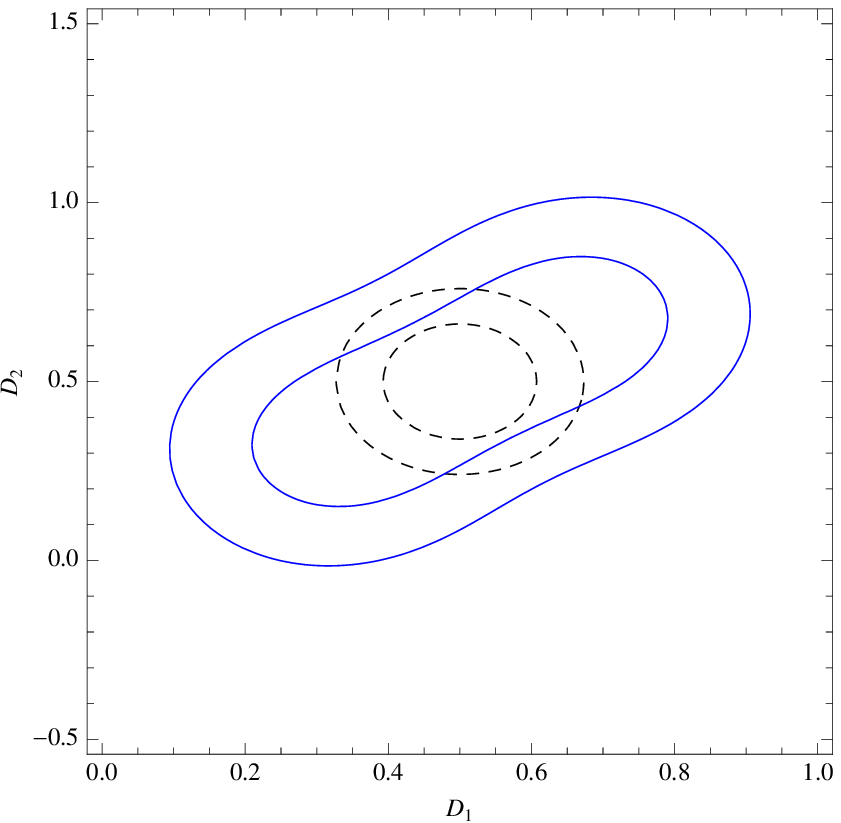}
 \caption{A toy example of how finite detection significance can affect even 1$\sigma$ combined constraints:  On the left, the solid (dashed) contours show the non-gaussian (gaussian) 1 and 2$\sigma$ constraints based on two measurements of $D_1$ and $D_2$, which each have 3.6$\sigma$ significance. The right panel shows the combined constraints, which are radically different with and without the assumption of gaussianity. }
 \label{BAO_comb}
\end{figure*}

We can go further and show how the non-gaussian sting can radically change even  $1-\sigma$ errors when combining constraints from different measurements. In Fig. (\ref{BAO_comb}) we show a toy model with two independent measurements of two  parameters $(D_1,D_2) = (0.3,0.3) \pm (0.1,0.15)$ and $(D_1,D_2)=(0.7,0.7)\pm (0.1,0.15)$, which are different due to some unaccounted for systematic error. Furthermore, each measurement is based on a 3.6$\sigma$ detection, which causes a slight expansion of 1 and 2$\sigma$ constraints (blue solid contours, corresponding to $\Delta\chi^2=$ 2.3 and 6) with respect to the gaussian constraints (black dotted contours), as described in this {\it letter} (see Equation \ref{dchi} below). However, as we see on the second panel, combining the two constraints leads to radically different posteriors,  with and without the assumption of gaussianity, \ie even the non-gaussian 1$\sigma$ combined errors are much bigger than the gaussian ones, illustrating that the effects need not only limited be limited to $>3\sigma$ results.

\begin{figure*}
\includegraphics[width=0.47\linewidth]{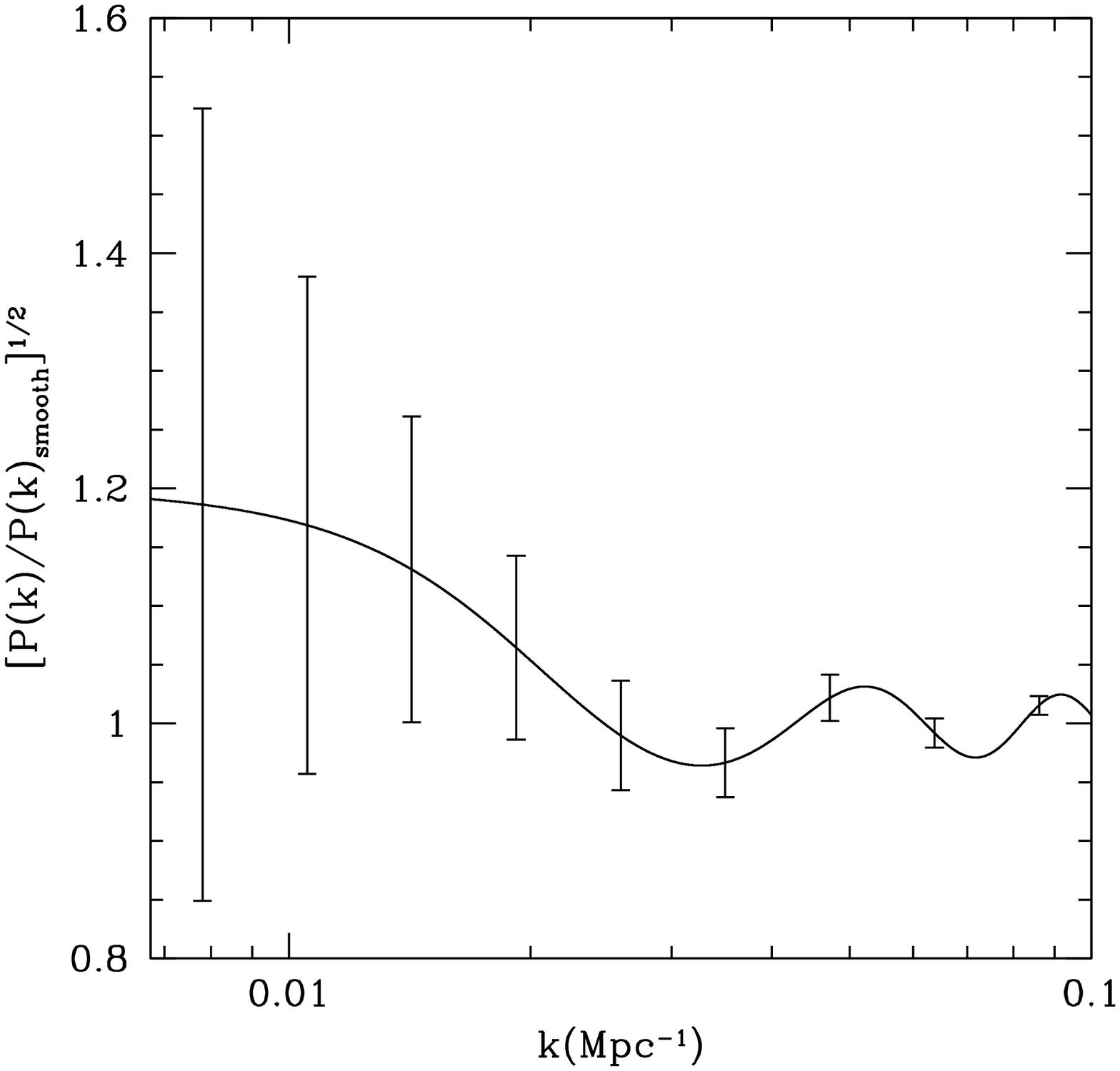}
\includegraphics[width=0.47\linewidth]{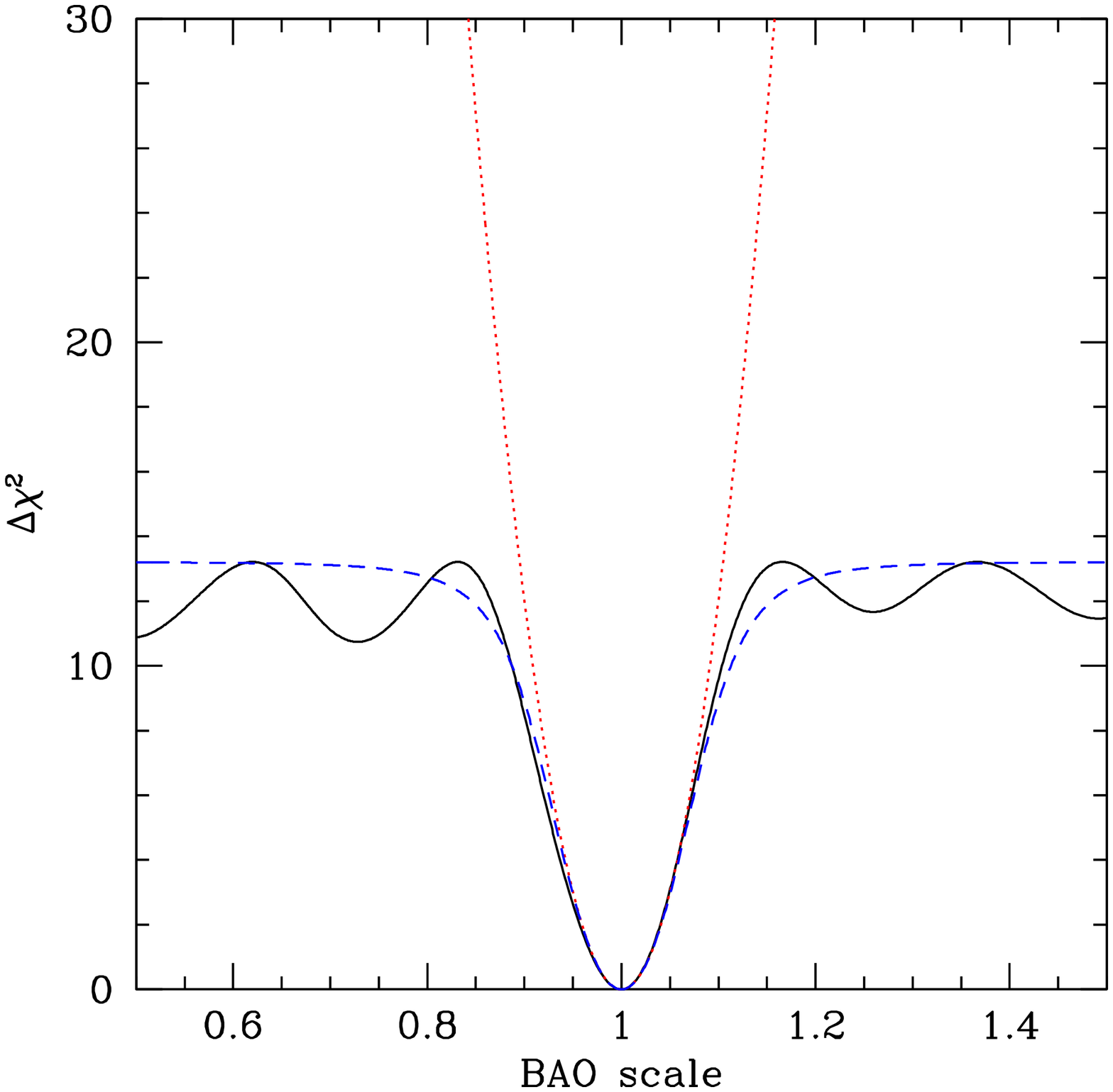}
 \caption{{\it Left:} (a)  Power spectrum of a hypothetical galaxy survey, normalized to a smooth power spectrum without BAO oscillations (see text for details). {\it Right:} (b) $\Delta\chi^2$ (= -2 log[likelihood]) for the position of the BAO peak in the linear power spectrum ($k<0.1~ {\rm Mpc}^{-1}$) for this survey. The red dotted  curve shows the quadratic/gaussian approximation which grows indefinitely, while the actual $\Delta\chi^2$ (black solid curve) saturates at $\sim (S/N)^2 = 13.1$ for the detection. The blue dashed curve is an interpolation between the gaussian and the asymptotic regime that we suggest in Eq. (\ref{dchi}) and is a good fit to the actual $\Delta\chi^2$.}
 \label{BAO_chi2}
\end{figure*}

The frequentist analog of this statement is that, for a finite detection confidence level, the function $\chi^2(\theta | d)$ should asymptote to a plateau far away from its minimum, rather than growing indefinitely. The maximum difference between the plateau and the minimum is roughly the square of the signal-to-noise of the detection, $S/N$. Therefore, the gaussian (quadratic) approximation to the likelihood ($\chi^2$) breaks down when $\Delta\chi^2_{\rm gauss} \gtrsim (S/N)^2$.

As an example, in Figure (\ref{BAO_chi2}) we consider the expected constraints on the BAO scale from a hypothetical galaxy survey with $10^6$ galaxies and bias of 1.5,  spread over a volume of 0.8 Gpc$^3$ at small redshifts \footnote{For this example, we use WMAP7+BAO+$H_0$ cosmology \citep{Komatsu:2010fb}.}. In Figure (\ref{BAO_chi2}a), we show the expected power spectrum+errors, normalized to a smooth power spectrum without BAO oscillations, using the  \citet{Eisenstein:1997jh} fit to the transfer function. The solid curve in Figure (\ref{BAO_chi2}b) shows $\Delta\chi^2$ for fitting this spectrum with a different BAO (or sound horizon) scale, marginalizing over the amplitude of the oscillations.     
Here, we only include linear scales, conservatively defined as $k<0.1 ~{\rm Mpc}^{-1}$, and assume gaussian errors for $\Delta(k) = \left[k^3P(k)/(2\pi^2)\right]^{1/2}$. 

While the gaussian approximation to the likelihood (dashed curve in Figure \ref{BAO_chi2}b), is a good approximation close to the minimum, it grows indefinitely, while the real $\Delta\chi^2$ saturates at  $\sim (S/N)^2$.  In order to reflect this, we propose a simple analytic function to approximate the true difference between $\chi^2$ and its minimum value:
\beq
\Delta\chi^2 \approx \frac{\Delta\chi^2_{\rm gauss} }{\left[1+(S/N)^{-4}\Delta\chi^4_{\rm gauss} \right]^{1/2}}.\label{dchi}
\eeq
As shown in Figure (\ref{BAO_chi2}b), this takes the quadratic shape of the gaussian approximation close to the minimum since the denominator is then negligible, but Eq. (\ref{dchi}) guarantees that $\Delta\chi^2$ remains smaller than $(S/N)^2$ of the detection, far from its minimum, which limits the statistical power of low signal-to-noise detections in constraining parameters. While the interpolating function (\ref{dchi}) is in good agreement with the actual $\Delta\chi^2$, we should note that it is only an approximation, and ideally one should use the full likelihood of the model fitting the data (\eg the full galaxy power spectrum) for an accurate statistical analysis.

\begin{figure}
\includegraphics[width=0.75\linewidth]{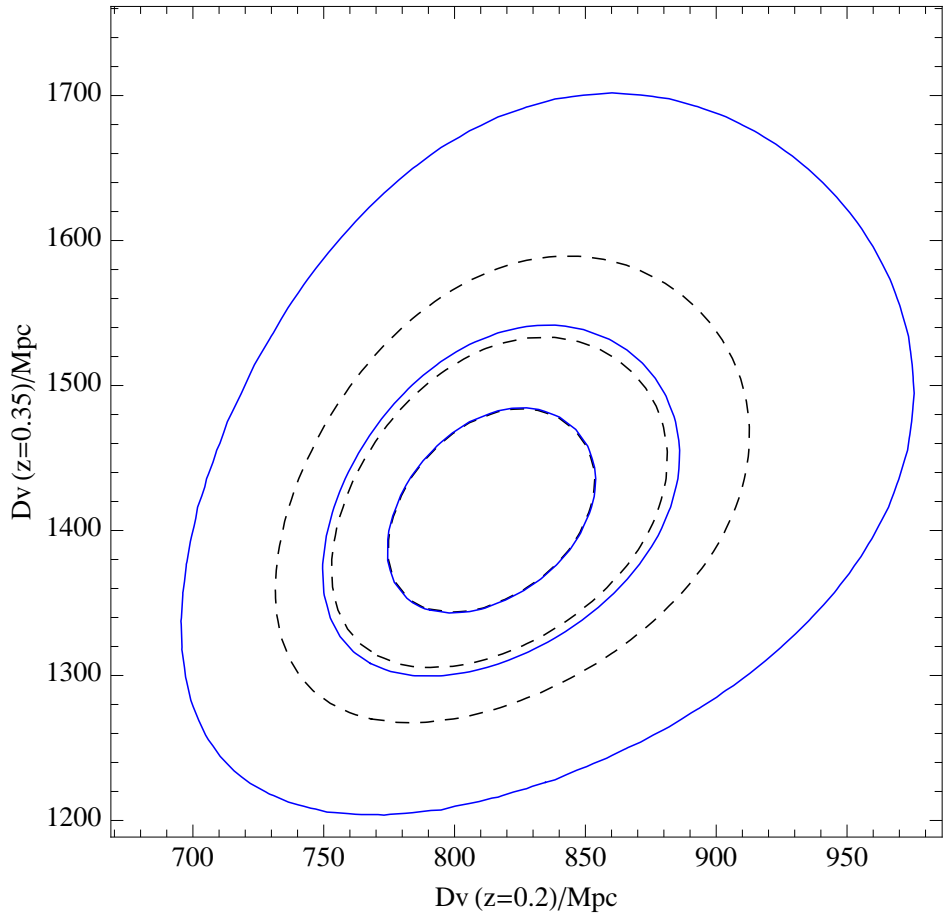}
 \caption{Contours of $\Delta\chi^2$ = -2 log[likelihood] = 2.3, 6.0, and 11.8 (corresponding to 68.3\%, 95.5\%, and 99.7\% gaussian probabilities) for distances to $z=0.2$ and $0.35$, from BAO detection in SDSS7 galaxies \citep{dr7}. The dashed black contours are gaussian fits quoted in \citet{dr7}, while solid blue contours correspond to our non-gaussian expectation, based on Eq. (\ref{dchi}) (compare to Figure 4 in \citet{dr7}). Although the 68.3\% (1$\sigma$) and 95.5\% (2$\sigma$) confidence contours are essentially unchanged, the effect of the finite probability of having detected the BAO peak has a profound effect on the 99.7\% (3$\sigma$) contour.}
 \label{BAO_percival}
\end{figure}

Finally, Figure (\ref{BAO_percival}) demonstrates the effect of non-gaussian posteriors on cosmological constraints. Here, we compare the gaussian approximation to likelihood distribution for distances to $z=0.2$ and $z=0.35$ in Sloan Digital Sky Survey \citep[SDSS;][]{dr7}, with our expectation from Equation (\ref{dchi}). Given that total $S/N$ for BAO detection in \citet{dr7} is $\sqrt{13.1}$ for two degrees of freedom, we see significant deviations from gaussian likelihoods beyond $99\%$ confidence level \footnote{Figure  (\ref{BAO_percival})  can be directly compared to Figure 4 in \citet{dr7}.  However, we should point out that \citet{dr7} underestimate the tail of the likelihood, as they ignore the (small) possibility of misidentifying the BAO peak in their mock realizations.}.

{\em Conclusions} -- In modern cosmology, as in many other
branches of science, there is a clear division of labour between
theorists and experimentalists. The theorists often desire a
simple and convenient data product that can be expressed in 
Gaussian form as a mean plus an error bar without using the full likelihood or redoing the initial analysis. This is exemplified by the many studies using 
the Gaussian approximation to the SDSS Baryon Acoustic
Oscillation (BAO) distance.

In the case of marginal detections, such as in the case of BAO cosmology or gravitational wave astronomy in the
near future, this will not suffice, and the possibility that the
detections are pure noise, even if small,  have to be taken
into account to get both precise and accurate results. The main
impact of taking this into account, is to radically alter the
relationship between confidence intervals. While $68\%$
confidence intervals may be essentially unchanged, $95\%$,
$99.7\%$ and higher confidence intervals can be radically
altered. In the Gaussian approximation, these intervals are all
trivially related to each other, but in the case where the
finite detection probability is included this relationship is
broken. Given that confidence limits must exceed $3\sigma$ to be
generally accepted as providing truly secure knowledge about
Nature, it is important to include finite detection
probabilities in cosmological statistical analyses lest one's
posterior is stung by the tail of the scorpion. We provide an
explicit and simple way to do this via Eq. (\ref{dchi}), which
uses the signal-to-noise of the detection to remove this sting.

A further concern might be that BAO analyses must assume a redshift-distance relation (typically $\Lambda$CDM) to compute the two-point correlation function in the first place, instead of recomputing $\xi(r)$ at every point in the cosmic parameter space. For models close to $\Lambda$CDM, this only causes small errors. However, if one wants to compute 3 - 5$\sigma$ constraints one now includes models - because of the detection effect discussed in this {\it letter}- that are far from $\Lambda$CDM and where the approximation of using $\Lambda$CDM to convert from redshift to distance might be very poor. This raises additional issues about the true constraints on cosmic acceleration from current BAO surveys.

Finally, it is a well-known phenomenon \citep[\eg][]{henrion} that experimentalists in all fields tend to underestimate the size of their error bars. This is particularly important here. Not only do the smaller error bars lead one to be over-confident about standard gaussian results, they are also likely to make one ignore the need for considering the finite-detection effects discussed here. For example, if one believes one has a 5 sigma detection of the BAO, then one could safely ignore the finite detection non-gaussianity. However, if one's error bars were too small by a factor of two that would make the non-gaussian corrections much larger. In the case of BAO, estimating error bars on the correlation function is very difficult. Ideally one should run a very large number of mock simulations \citep[\eg][]{baomocks}, but in the absence of this it is common to use approximations such as jack-knife or log-normal simulations which may not give the correct answer. 

The simplest way to take care of all of this effect is to use the full $P(k)$ power spectrum measurements in Monte-Carlo Markov Chains, as is possible in \eg CosmoMC.

{\em Acknowledgements:} We thank Martin Kunz, Ren\'{e}e Hlozek, Hiranya Peiris and Will Percival for insights and particularly Chris Blake, who has always demanded $5\sigma$ for BAO detections. NA and BB are partially supported by the Perimeter Institute (PI). Research at PI is
supported by the Government of Canada through Industry Canada and by the
Province of Ontario through the Ministry of Research \& Innovation. BB is supported by the South African National Research Foundation.



\end{document}